\documentstyle[preprint,balanced,epsfig,prl,aps]{revtex}
\begin{document}
\draft
 
\title{Learning from mistakes} 
\author{Dante R. Chialvo*\cite{drcemail} and Per Bak*\cite{bakemail}}
\address{{$\ddagger$}Niels Bohr Institute, Blegdamsvej 17, Copenhagen, Denmark.  
{$\dagger$}Division of Neural Systems, Memory and Aging, University of  Arizona, Tucson, AZ 85724, USA. 
{*}Santa Fe Institute, 
 1399 Hyde Park Rd., Santa Fe, NM 87501, USA.
\\
}
\date{\today}
\maketitle

\begin{abstract}
A simple model of self-organised learning with no classical 
(Hebbian) reinforcement is presented. Synaptic connections involved in 
mistakes are depressed. The model operates at a highly adaptive, probably 
critical, state reached by extremal dynamics similar to that of recent 
evolution models. Thus, one might think of the mechanism as synaptic 
Darwinism.
\end{abstract} 
    
\vspace{1.truein} 

It is widely believed that learning in the brain resides in alterations 
of synaptic efficacy. Without exception, contemporary formulations of such 
learning follows Hebb's ideas \cite{bible} of reinforcement: synaptic 
connections among neurons excited during a a given firing pattern are 
strengthened by a process  of long term potentiation (LTP). 

However, long term synaptic depression (LTD) in the mammalian brain is almost 
as prevalent as potentiation, but there appears to be little or no
understanding of its functional role. 
Working hypotheses covers a wide range, where depression is given 
always an auxiliary function to potentiation \cite{weak}.
A recent review \cite{barnes}, reflecting the current variety of ideas 
regarding the functional role of LTD, speculates:
``Although it is conceivable that LTP is the critical phenomena used for
storing information, and that LTD may exist simply to reset LTP, it must be
noted that it is also conceivable for the converse to be true.''
 
We present an alternative to Hebbian learning. Turning things upside down, we 
suggest that LTD is, in some instances of learning and development, the fundamental
mechanism with LTP playing a secondary role. This view is supported by
studies of a simple neuronal learning model. There are two fundamental
differences between the classical view of learning by reinforcement and
the view discussed here:
 
1) Learning by reinforcing good responses is a process that by definition 
never stops. There is not an explicit rule that ends the reinforcement 
whenever the goal has been reached. On the other hand, if learning proceeds only 
by correcting mistakes it implies a process that stops as soon as the goal is 
achieved. This prevents formation of ``deep holes'', i.e. highly stable 
states from which adaptation to new rules is difficult and slow, requiring,
perhaps, a significant amount of random noise.

2) If an adaptive system is placed on a new environment, or otherwise
subjected to learn something new, the likelihood of making mistakes 
is generally larger than the chance to be initially right.  
Therefore, the opportunity to shape synapses is larger for the  
adaptive mechanism that only relies on mistakes, leading to faster
convergence. 

In order to develop these ideas, a model of a adaptive neural structure 
has been constructed.   
Although it is only a caricature of a real brain,
all the ingredients are biologically reasonable and correspond to well-
documented physiological processes. The model is 
completely self-organised with no need for external computation of
synaptic strengths, in contrast to, for instance, feed-forward and 
back-propagation neural networks. All control mechanisms are local at
the post-synaptic site of the active neurons, and information 
is fed back globally to all neurons.

Each neuron receives input from, and sends output to, several other neurons. 
Just about any arbitrary architecture, for instance a completely random one, 
can be chosen. For descriptive purposes, however, consider a two layer 
network where $K$ represents the outputs, $I$ the inputs and $J$ the middle layer 
(Figure 1). Each input is connected with each neuron in the middle layer which,
in turn, is connected with each output neuron, with weights $W$ representing
the synaptic strengths. 
The network must learn to connect each input with 
the proper output for any arbitrary associative mapping. The weights are initially randomised, $0<W<1$.

In order to achieve efficient self-organised learning, it is essential to
keep the activity low \cite {stassi}. Here, we assume that {\it only one}
neuron $k$, namely the one which has the largest $w(k,j)$, fires at each
time step \cite{wta}. This type of ``extremal dynamics'' is known to organise
dynamical systems into a highly adaptive (high susceptibility) critical state,
\cite {soc1} \cite{soc2} most notably in recent models of biological 
evolution \cite{BS}.

The dynamical process in its entirety is as follows:\\
An input neuron is chosen. The neuron $j_m$ in the middle layer with the 
largest $w(j,i)$ is firing. Next, the output neuron $k_m$ with the maximum 
$w(k,j_m)$ is firing. If the output $k$ happens
to be the desired one, {\bf  nothing }is done, otherwise
$w(k_m,j_m)$ and $w(j_m,i)$ are both depressed by a fixed amount $\delta$,
which is redistributed among the other incoming synapses  to the same two 
neurons. The redistribution can be either uniform, or to one randomly 
selected input.

The iterative application of this rule leads to a quick convergence to any 
arbitrary input-output mapping. Figure 2 shows this for a
map (labelled ``a") where seven inputs 1-7 are mapped to the corresponding
seven outputs, 1-7, in a few hundred time steps.
Re-mapping to new cases is straightforward for this
network. After learning the 
identity map  the network is exposed successively to
five other associative maps, labelled ``b" through ``e" respectively,
where the definition of correct outputs have been modified. It can be seen
that the error (plotted in the top diagram) quickly returns to zero indicating
that the new pattern has been completely learned.

The reason for quick re-learning (adaptation)
is simple. The rule of adaptation assures that synaptic changes only
occur at neurons involved in wrong outputs. The landscape
of weights is only re-shaped to the point where the new winners barely 
support the new correct output, with the old pattern only slightly
suppressed. Thus, only a slight suppression of a currently active pattern is
needed in order to generate new patterns when need be. In particular, 
re-learning of ``old'' patterns which have been correct once in the past is
fast.  This is illustrated in 
figure 2, where the pattern ``a" is learned much faster the second time. The 
landscape of synaptic strength in our model after many learning cycles 
consist of very many values  which are very close to those of the
active ones, a manifestation of the critical nature of the state.

This contrasts with the classical reinforcement scenario where 
at the end of some pre-established learning period the correct synapses
are dominating the incorrect ones. Adaptation is slow, i.e. 
new learning has to
start essentially from scratch, and there is  no memory kept of old patterns.

The scaling of the learning time with the size of the
middle layer is interesting. A network with a large middle layer offers 
many more
options to the system to select amongst, when an incorrect pattern is
suppressed. This  speeds up the learning of the correct
associations, as shown in Figure 3 where
the average learning time is plotted for a network with
constant number (seven used here) of units in the $I$ and $K$ layers and increasing
number of neurons in the middle layer. It can be seen that 
performance improves for increasing ratios, an increase of one
order of magnitude in the divergence ratio speed up learning two orders of
magnitude. 
Results for re-learning scale in the same qualitative way. Bigger is
better, in contrast with reinforcement methods where learning is slower in 
bigger systems. For constant middle layer' size, the scaling of learning time versus input (and output) layer
size $N$  goes $\sim$ $N^2$, also in contrast with reinforcement models where
the scaling is usually exponential on $N$.

In order to illustrate the robustness of the model, which is important for
our mechanism to have any biological relevance, we have studied an architecture
where each of $n$ neuron is arbitrarily connected to a number $n_c$ of other 
neurons. A number of neurons ($n_i$ and $n_o$) are arbitrarily selected as 
input and output neurons, respectively.
If, after a number of firing events, the correct output has not been reached,
each synapse in the chain of firings is depressed as before. If the correct
output is achieved there is no modification. A system with $n=25$, 
$n_i=n_c=n_o=5$ behaves like the layered structure presented above. This
illustrates the development of structure
even in the case where  all initial connections are absolutely uncorrelated.
 
In addition to giving insight into mechanisms for learning in the brain,
the ideas presented here could be useful for other artificial learning 
processes. To demonstrate this point an ``agent'' has been constructed using 
a two layers architecture (similar to the one discussed above). The agent is 
supposed to track a moving target, which makes one (or more) steps randomly
to the left 
or to the right at each time step. The ``sensory'' input layer gives the 
position of the agent relative to the target. The output cells can be thought 
of as muscles moving the agent to the left or to the right by various amounts.
A correct output  is one which takes the agent closer to the target, a wrong 
output, triggering modifications of synaptic strengths, is one that does
not. Seven inputs and seven outputs, and three hundred units in the middle 
layer were used.  Figure 4 illustrates the initial process of learning, 
and the response to various drastic perturbations to the network\cite{code}
The network is able to discover all by itself the proper map 
of connections between sensory and motor neurons that ensures a perfect 
tracking.

The biological plausibility of the schema depends on the realization at 
the neuronal level of two crucial features:

a) Activity propagates through the strongest connections, i.e. a winner-takes-all 
action. This can be fulfilled by a local circuit organisation, known to exist in all
cortices, where
feed-forward and recurrent excitations competes with lateral and feedback
inhibitory connections. The robustness of such operation have been extensively
studied on detailed neuronal models \cite{koho}. The coexistence of LTD in 
some of these structures might not be just coincidental \cite{LTDcere}.
Alternatively, a global threshold mechanism keeping the firing rate low would
suffice.

b) Depression of synaptic efficacy involves the entire path of firing
neurons. A process must exist such that punishment can be relayed long after
the neuron has fired, when the response from the outer world to the action
is known. We conjecture a mechanism of ``tagging''  synapses 
for subsequent LTD,  analogous to (but mirroring) recently 
reported tagging of synapses for LTP \cite{tagging}.

Historically, processes that were thought of as directed learning have 
been shown to be caused by selection. The Larmarquean view of 
evolution as a learning process, where useful acquired features are
strengthened was replaced by the Darwinian view of evolution as a selection 
process, where the unfit species are weeded out. A similar paradigm shift 
occurred in immunology. Ironically, if our thinking turns out to be correct, 
learning is not a (directed) learning process either, but also an evolutionary
selection process where incorrect connections are weakened.

\pagebreak

\begin{figure}[htbp]
\centering
\psfig{figure=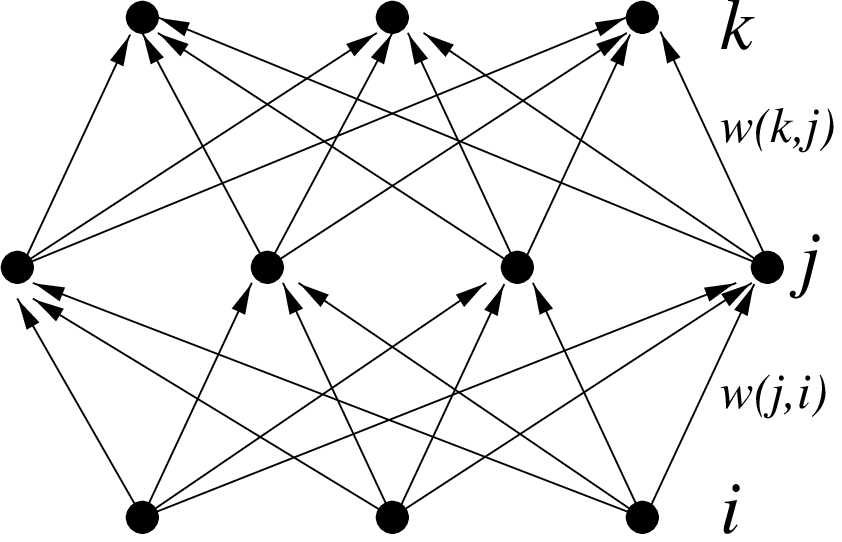,width=2.5truein,angle=-90}
\vspace{1.truein}
\caption{\footnotesize{Neurons in layer $J$ have synaptic connections with all
inputs in $I$ and connect with all neurons in layer $K$.} }
\end{figure}

\pagebreak
\begin{figure}[htbp]
\centering
\psfig{figure=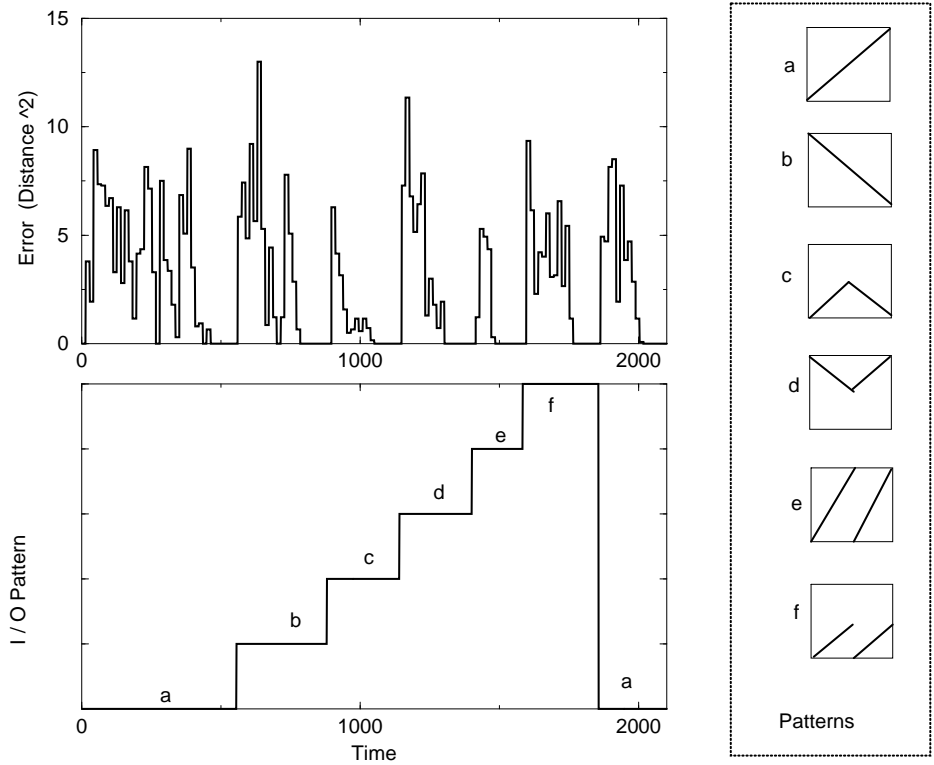,width=4.truein,angle=-90}
\vspace{1.truein}
\caption{
\footnotesize{Learning six associative mapping (patterns illustrated in the
right box, showing the outputs (y-axis) to be connected with the inputs
(x-axis). Seven inputs and outputs and three hundreds neurons in the
middle layer were used with $\delta =.01$. The bottom diagram shows which of the six
patterns is being
presented at a given time. The top diagram shows the relative distance
between the desired output and the net's current output.
A distance of zero means perfect learning has been achieved;
after 50 additional iterations, (for visualization purposes),
 a new pattern is presented} }
\end{figure}
\pagebreak
\begin{figure}[htbp]
\vspace{-.5truein}
\centering
\psfig{figure=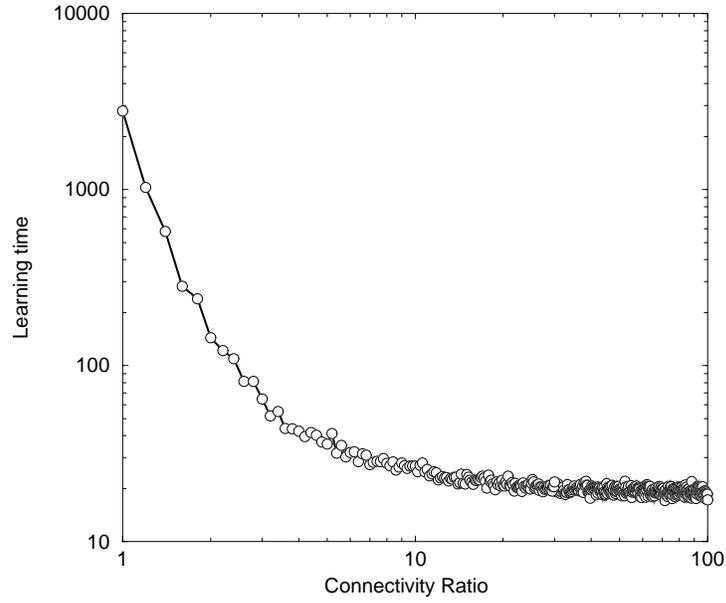,width=4.truein,angle=-90,clip=true}
\vspace{1.truein}
\caption{\footnotesize{Scaling of the average learning time  (per input)
as a function of the middle layer/input layer ratio. Values represent means of
128 realizations. Learning time is defined as the number of steps required to learn all 
input-output associations.}}
\end{figure}

\pagebreak
\begin{figure}[htbp]
\centering
\psfig{figure=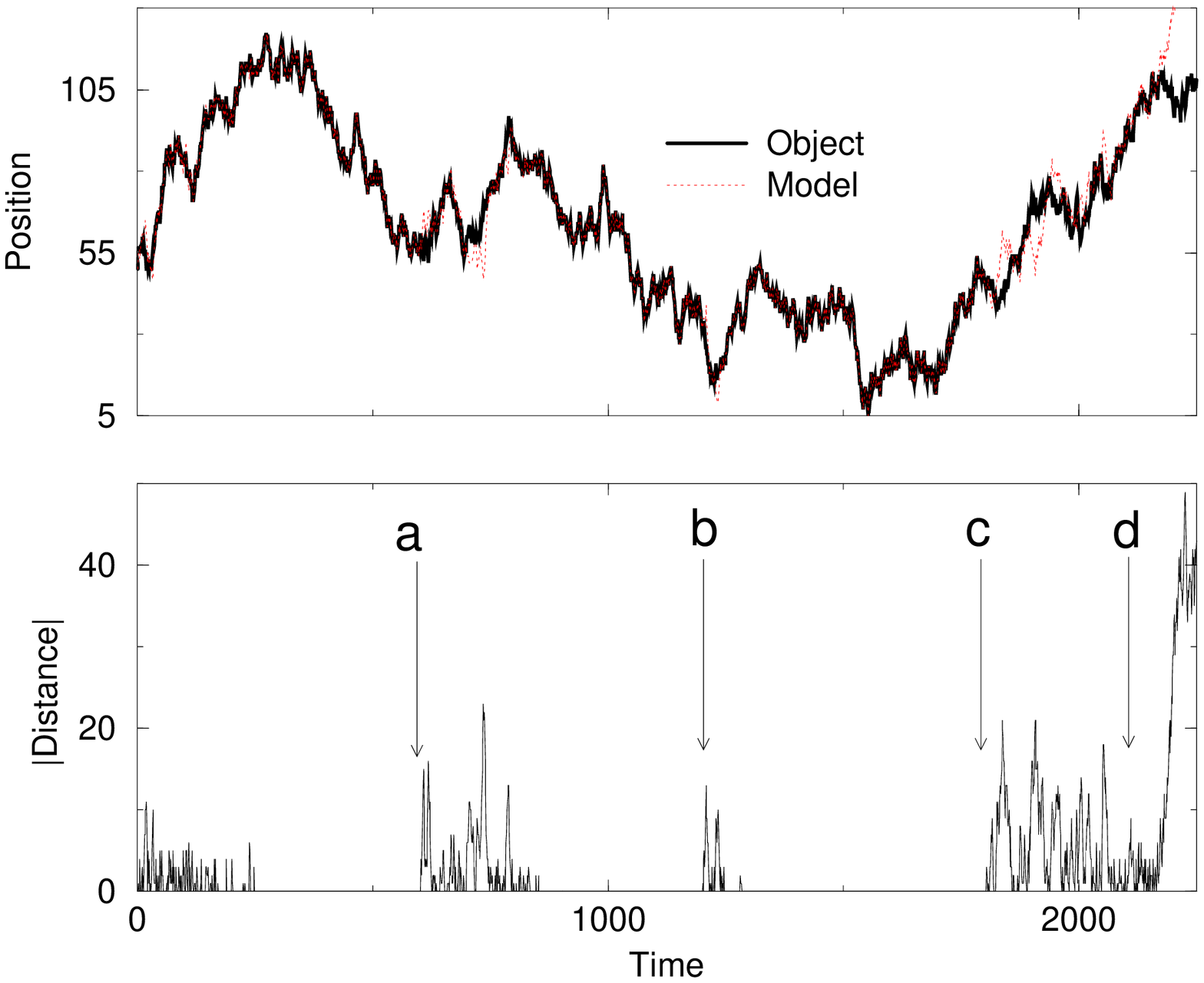,width=5.truein,angle=-90}
\vspace{.5truein}
\caption{
\footnotesize{An autonomous agent learning to track a
randomly moving object, subjected to
various (some drastic) perturbations.
Top diagram: Position of the object (continuous line) and of the agent
(dashed line) as a function of time. Bottom diagram: Tracking error as a
function of time
expressed as absolute value of the distance between the object and the model.
At about 250 time steps the model has learned to perfectly track the object.{\bf(a):} At the
time indicated by the arrow the inputs are inverted. The network re-maps
 in s few hundred  steps
and the error goes back to zero.
{\bf(b):} 75 per cent of the synapses are randomised at the time indicated by
the arrow. The network quickly reorganises and decreases the error.
{\bf(c):} 75 per cent of the synapses are removed from the middle layer.
The error grows and although it returns to relatively small values, it is no
longer able to perfectly track the object.
{\bf(d):} 75 per cent of the input synapses are removed. The network
can not cope with the damage and the agent takes off in the wrong direction.
 } }
\end{figure}


\begin{references}
\bibitem {weak} Implicit or explicit inclusion of some sort of depression
is most commonly included in models for mathematical
convenience in order to prevent unbounded growth of the synaptic weights. 
Is is mostly justified by the mathematics of the models rather by the
biology supposed to be represented in them.
\end{references}
\end{document}